\documentclass[journal,twoside,web]{ieeecolor}

\usepackage{generic}
\usepackage{tcolorbox}
\usepackage{cite}
\usepackage{mathrsfs}
\usepackage{comment}
\usepackage{amsmath,amssymb,amsfonts}
\usepackage{graphicx}
\usepackage[unicode,psdextra]{hyperref} 
\usepackage{bookmark}                    
\usepackage{mdframed}

\newcommand{\comb}{\mathrm{III}}
\newcommand{\ztransform}{\mathcal{Z}}
\newcommand{\laplace}{\mathscr{L}}

\newcommand{\res}{\mathrm{Res}}
\newcommand{\arcR}{\mathrm{Arc}_R}

\newcommand{\Ts}{T_s}
\newcommand{\ws}{\omega_s}

\usepackage{mathtools}
\mathtoolsset{centercolon}

\usepackage{mathrsfs}
\usepackage{mathtools}
\usepackage{url}
\usepackage{cite}
\usepackage{booktabs}

\usepackage{hyperref}
\hypersetup{pdfborder={0 0 1}, colorlinks=false, citecolor=blue, linkbordercolor=red }
\usepackage{xcolor}
\usepackage{empheq}
\usepackage{physics}  
\usepackage{float}
\newtheorem{theorem}{Theorem}
\usepackage{algorithm,algorithmic}
\usepackage{textcomp}

\DeclareMathAlphabet{\mathpzc}{OT1}{pzc}{m}{it}

\newtheorem{lemma}[theorem]{Lemma}

\def\BibTeX{{\rm B\kern-.05em{\sc i\kern-.025em b}\kern-.08em
    T\kern-.1667em\lower.7ex\hbox{E}\kern-.125emX}}
\begin{document}
\title{Foundational Correction of $\ztransform$-Transform Theory: Restoring Mathematical Completeness in Sampled-Data Systems}

\author{Yuxin Yang, \IEEEmembership{Member, IEEE}, Hang Zhou \IEEEmembership{Member, IEEE}, Chaojie Li, \IEEEmembership{Member, IEEE}, Xin Li,\IEEEmembership{Member, IEEE},Yingyi Yan,\IEEEmembership{Member, IEEE},  Mingyang Zheng, 
\thanks{This work is intend to be submitted to IEEE transactions on automatic control }
\thanks{Yuxin Yang, Hang Zhou and Chaojie Li are with the School of Electrical Engineering \& Telecommunication University of New South Wales, CO 80305 Australia (e-mail: Yuxin Yang: z5307358@ad.unsw.edu.au; Hang Zhou: z5018464@ad.unsw.edu.au; chaojie.li: cjlee.cqu@163.com;). }
\thanks{Xin Li is with the School of Electrical Engineering University of Southeast, 210096 Sipai Road, Xuanwu District, China (e-mail: li-xin@seu.edu.cn). }
\thanks{Yingyi Yan is with 
the School of Integrated Circuit Science and Engineering, UESTC, Chengdu, China (e-mail: author@nrim.go.jp).}
\thanks{Mingyang Zheng is with Incosync Limited corporation (e-mail: 723461020@qq.com).}
}

\maketitle

\begin{abstract}
This paper identifies and rectifies a fundamental mathematical flaw in the classical formulation of the $\ztransform$-transform and its relationship to the inverse Laplace transform ($\laplace^{-1}$). We demonstrate that conventional implementations—based on residue calculus in classical control and Dunford-Taylor integrals in modern state-space methods—systematically neglect the contribution from the infinite arc in the complex plane. This oversight induces discontinuity errors at critical points (e.g., $t=0$) and propagates inconsistencies into step-function definitions. By incorporating the full Bromwich contour integral with all boundary contributions, we establish a mathematically complete framework that restores consistency between $\laplace^{-1}$, $\ztransform$-transform, and DTFT aliasing theory. Our correction resolves decades of contention regarding initial-value discrepancies and extends to modified $\ztransform$-transforms for delayed systems. The work transitions sampled-data theory from empirical patching to mathematical completeness, providing rigorous foundations for both classical and modern control applications.
\end{abstract}
\begin{IEEEkeywords}
Z-transform, inverse Laplace transform, sampled-data systems, Bromwich integral, residue theorem, Dunford-Taylor integral, discrete-time systems
\end{IEEEkeywords}

\section{Introduction}
\IEEEPARstart{M}{athematical} foundations of sampled-data theory have long relied on two flawed implementations of the inverse Laplace transform ($\laplace^{-1}$): \textbf{residue calculus in classical control} and \textbf{Dunford-Taylor integrals in modern state-space methods}. This paper reveals how both approaches systematically neglect critical contributions from the infinite arc in the complex plane—a fundamental oversight that propagates errors into $\ztransform$-transform theory, step-function definitions, and physical system modeling.  

\subsection{The Dual Pathology of Inverse Laplace Implementations}

\subsubsection{Classical Control: Misapplied Residue Calculus}
Traditional $\mathcal{Z}$ transform methods compute sampled sequences as:
\begin{equation}
x_c(n\Ts) = \sum \res\left[X_c(s)e^{sn\Ts}\right]
\label{eq:residue_error}
\end{equation}
This erroneously assumes $\int_{\arcR} \!\!\! X_c(s)e^{st}ds \to 0$—valid for $t>0$ but \textbf{indefensible at $t=0$}:
\begin{IEEEeqnarray}{c}
\lim_{R\to\infty} \int_{\arcR} \!\!\! X_c(s)e^{s\cdot0}ds \neq 0 
\label{eq:arc_error}
\end{IEEEeqnarray}
The neglected term introduces \textbf{discontinuity errors at initialization}.

\subsubsection{Modern Control: Misapplied Dunford–Taylor Integrals in Modern Control}
Traditional State-space discretizations and inverse Laplace transform for the state space resolvent $(sI-A)^{-1}$ employ:
\begin{equation}
    \begin{aligned}
    &\frac{1}{2\pi i}
    \oint_{\Gamma}\!e^{s\,t}\,\bigl(sI - A\bigr)^{-1}\,ds\;=\;
    e^{A\,t}\\ 
    \label{eq:dunford_error}
    \end{aligned}
\end{equation}
where $\Gamma$ is the Bromwich integral path for the Inverse Laplace transform.
In modern control theory, all the $t$ values, including $t=0$ , equation \eqref{eq:dunford_error} is applicable.
This paper proved that equation \eqref{eq:dunford_error} is only applicable when $t \neq 0$. For $t=0$ case, the following equation holds instead:
\begin{align}
&\frac{1}{2\pi i}\int_{\Gamma}(s I - A)^{-1}\,\mathrm{d}s
=\tfrac12I
\end{align}
\subsection{Ambiguous Step-Function Definitions}
For the causal transfer function, the inverse Laplace transform inherits the Heaviside function.
However, heaviside $u(t)$ suffers inconsistent definitions:
\begin{equation}
    \begin{aligned}
    &\text{Left-continuous: } u(0)=0 \\
    &\text{Right-continuous: } u(0)=1 \\
    &\text{Algebraic-mean: } u(0)=0.5
    \end{aligned}
\end{equation}
Sampled-data theory defaults to \textbf{right-limit convention }. This paper has rigorously proved that the Algebraic-mean should be utilized.
\subsection{Contradiction between conventional $\mathcal{L}^{-1}$ and DTFT}
Traditional DTFT proposed that 
\begin{equation}
X_s(z) = \frac{1}{\Ts}\sum_{k=-\infty}^{\infty}X\left(s - j\tfrac{2\pi}{\Ts}k\right)
\label{eq:definition_fix}
\end{equation}
\subsection{Superficial Fixes and Their Failures}

\subsubsection{Definition-Centric Correction}
\begin{equation}
X_s(z) = \frac{1}{\Ts}\sum_{k=-\infty}^{\infty}X\left(s - j\tfrac{2\pi}{\Ts}k\right) + \frac{x(0^+)}{2}
\label{eq:definition_fix}
\end{equation}
\textbf{Flaws}:
\begin{itemize}
\item Inherits infinite-arc error
\item Enforce non-rigorous $u(0)=1$
\item Contradicts DTFT without any rigorous justification. except claiming that the right-continuous should be applied according to empirical engineering practice.
\end{itemize}

\subsubsection{Limit based Correction}
\begin{equation}
X_{\text{imp}}(z) = \frac{1}{\Ts} \sum_{k} X\left(s - j \tfrac{2\pi}{\Ts} k \right) - \frac{x(0^+)}{2}
\label{eq:method_fix}
\end{equation}
\textbf{Flaws}:
\begin{itemize}
\item Lacks mathematical proof
\item Enforces non-rigorous $u(0)=0$
\end{itemize}

\begin{center}
\fbox{\parbox{\columnwidth}{\centering \textbf{Core Issue}: Both fixes neglect the Big Arc Integral and step-function arbitrariness}}
\end{center}

\subsection{Foundational Correction}

\subsubsection{Rigorous Inverse Laplace via Full Bromwich Integration}
We rectify $\laplace^{-1}$:
\begin{equation}
\begin{aligned}
&\frac{1}{2\pi j}
\int_{c-j\infty}^{c+j\infty}
G(p)\exp(pt)\,dp\\
&=
\sum_{\Res\{p_k\}<c}
\Res_{p=p_k}G(p)\exp(pt)\\&-\frac{1}{2\pi j}
\int_{\arcR}^{}
G(p)\exp(pt)\,dp
\;
\end{aligned}
\end{equation}
\textbf{Key proof}: The 
\begin{equation}
\frac{1}{2\pi j}
\int_{\arcR}^{}
G(p)\exp(pt)\,dp = \frac{a_0}{2} \neq 0
\label{eq:arc_contribution} \textbf{when $t = 0$} 
\end{equation}
where, $a_0 = g(0^+)$ 
\subsubsection{Corrected Heaviside Step-Function Definition}
Sampled systems require:
\begin{equation}
u(0) = \frac{u(0^-) + u(0^+)}{2} = 0.5
\label{eq:step_correction}
\end{equation}

\subsection{Triple Unification}
    \begin{table}[H]
        \begin{tabular}{ll}
        \textbf{Triple Unification}\\
         
          \textbf{Domain} & \textbf{Consistency Restoration} \\
          \cline{1-2}
          $\laplace^{-1} \leftrightarrow \ztransform$ & sampled  Rigorous $\laplace^{-1}$ at $t=n\Ts$ $\equiv$ $\ztransform$ \\
          $\ztransform \leftrightarrow \text{DTFT}$ & Corrected $\ztransform = \frac{1}{\Ts}\sum_k X(s+jk\ws)$ \\
          Math $\leftrightarrow$ Heaviside step & $u(0)=0.5$ instead of $u(0)=1$\\
          \cline{1-2}\\
          \textbf{definition for $w_s$: $w_s =\frac{ 2\pi}{T_s}$}
        \end{tabular}
    \end{table}

\subsection{Engineering Impact}
\begin{enumerate}
\item \textbf{Classical}: Corrects residue-based methods
\item \textbf{Modern}: Fixes Dunford integrals to fix the flaws on the inverse laplace transform of state space resolvent
\end{enumerate}

\section{Literature Review}

The $\mathcal{Z}$-transform is commonly used to characterize the frequency-domain behavior of discrete-time signals and to describe the dynamic response of discrete-time difference equation systems\cite{Ragaztransform}. It is important to note that many applications of the $\mathcal{Z}$-transform arise in the context of sampled-data systems—hybrid systems that involve discrete sampling operations(ADCs) within continuous-time processes.

To develop a discrete-time $\mathcal{Z}$-domain model for such systems, we must model the behavior of the system following the sampling operation. This modeling process defines the sampled-data system, in which the $\mathcal{Z}$-domain representation is effectively derived from the original continuous-time system by means of an $\mathpzc{Laplace}$-to-$\mathcal{Z}$ mapping. This approach is often associated with the impulse response invariance method. In cases where the continuous-time system includes a sample-and-hold component before sampling, the modeling procedure is commonly referred to as zero-order hold (ZOH) discretization. Both methods are widely employed to derive discrete-time filters from continuous-time prototypes in digital signal processing applications. The ZOH discretization method admits a matrix-based formulation, which allows direct mapping from a continuous-time state-space model to its discrete-time form without requiring partial fraction expansion or table lookup. This matrix formulation was introduced by John R. Ragazzini and Gene F. Franklin in their co-authored book~\cite{Franklin}.  

In contrast, the impulse response invariance method does not have a matrix-based formulation in the existing literature. At present, the mathematical rigor of the $\mathcal{Z}$-transform is demonstrated by showing its equivalence to the DTFT. However, this equivalence has been questioned due to the main reason that the aliasing series sometimes does not match the entries in the $\mathcal{Z}$-transform table \cite{Jackson2000}, \cite{Wolfgang2000}. 
\[
X_s(s)\;=\;\frac{1}{T_s}\sum_{k=-\infty}^{\infty} X\bigl(s + j\,k\,\omega_s\bigr), \qquad \omega_s=\tfrac{2\pi}{T}.
\]

Since the book they referenced define the $\mathcal{Z}$ transform as:
\[
  X_s(z)
  = \frac{1}{T}\sum_{k=-\infty}^{\infty}X\!\bigl(s - j\tfrac{2\pi}{T}k\bigr).
  \;
\]

There are two different explanations for this issue in the academic community.

\subsection{Definition-Centric Correction}
\label{Defenitioncentric}
Some textbooks resolve the aliasing discrepancy by amending the $\mathcal{Z}$-transform definition itself\cite{zadeh1963linear},\cite{wilts1960}.  These works argue that, at a first-kind discontinuity, the right-hand limit of \(x(nT)\) must be used.  Mathematically this adds a half-sample term to the standard transform pair, restoring agreement with the aliasing series:
\[
  X_s(\mathcal{Z})
  = \frac{1}{T}\sum_{k=-\infty}^{\infty}X\!\bigl(s - j\tfrac{2\pi}{T}k\bigr)
  \;+\;\frac{x(0^+)}{2}.
\]
We call this the Definition-Centric Correction because it leaves the impulse-invariance method untouched and merely tweaks the tabulated $\mathcal{Z}$-transform entries.

The Definition-Centric Correction does not appear in the early influential papers on sampled-data theory~\cite{JohnAIEEtac},\cite{Jurypaper}. Instead, the correction was introduced in two textbooks published in the 1960s ~\cite{wilts1960},\cite{zadeh1963linear},\cite{Jury_zTransform}. We will show in \textbf{Section \ref{original_process}} that this correction is not mathematically rigorous. However, many textbooks do not include this correction at all. For example, in the mid-1970s, Jury included~\cite{wilts1960} in his new book~\cite{Jury_zTransform}. 

After this correction, engineers must add an initial-value term when using the $\mathcal{Z}$-transform table on a series derived from a continuous-time prototype. This change removes contradictions between the tabulated $\mathcal{Z}$-transform mappings and the aliasing series. \textcolor{red}{The result is mathematically consistent.} However, this correction relies solely on choosing the right-hand limit at a first-kind discontinuity. This choice matches common engineering definitions of the Heaviside step function and the inverse Laplace transform. However, it does not guarantee that the $\mathcal{Z}$-transform defined this way matches the physical result of convolving the impulse response with a Dirac comb. In \textbf{Section \ref{original_process}}, we will show that this rule does not reflect the true behavior of sampled-data systems. We will argue that values at a discontinuity should use the arithmetic mean of the left- and right-hand limits. \textcolor{red}{Similar issues due to step function definitions appear in modern control theory.} \textbf{Section \ref{Correction}} will present corrections for these cases from a functional analysis viewpoint.

\subsection{Method-Centric Alignment}
\label{methodcentric}
An alternative line of thought focuses on the sampling method itself, insisting that the impulse-invariance construction should exactly reproduce the aliasing series with changing the impulse-invariance-table\cite{Jackson2000}. 

\begin{align}
X_{\text{impulse\_invariance}}(\mathcal{Z})
&= \frac{1}{T} \sum_{k=-\infty}^{\infty} X\!\left(s - j \tfrac{2\pi}{T} k \right)  \\
&= X_s(\mathcal{Z}) - \frac{x(0^+)}{2}.
\end{align}

Proponents of this Method-Centric Alignment introduce an initial-value term into the impulse-invariance formula. Its applicability has been limited to filter design. It does not consider the possible impact of the same flaw on the modeling of sampled-data systems. It also overlooks the fact that this correction actually challenges the original definition of the $\mathcal{Z}$-transform. However, this approach lacks detailed mathematical justification. 

\subsection{On the influence of two views}

In current applications, two viewpoints coexist. The Definition-Centric Correction is little known. The Method-Centric Alignment does not question the $\mathcal{Z}$-transform definition itself. Although the Method-Centric Alignment has gained some influence, it remains confined to filter design methods.

Because the Definition-Centric Correction is rarely cited, many researchers claim that their impulse-invariance models produce the aliasing series and match the $\mathcal{Z}$-transform table without any extra term~\cite{xinli2018}. This leads to results that are mathematically incorrect and internally inconsistent.

Some scholars later recognized inaccuracy result using the $\mathcal{Z}$-transform approach and adopted pure mathematical techniques to evaluate the series ~\cite{YanNaPartI},\cite{ztransformpll}. However, they did not question the underlying definition of the $\mathcal{Z}$-transform. The influence of the $\mathcal{Z}$-transform in sampled-data modeling comes largely from its simplicity in replacing series summation. The shift to direct series evaluation reveals the limits of the $\mathcal{Z}$-transform framework in this context.

Therefore, clarifying the correct definition of the $\mathcal{Z}$-transform and its correspondence to both series summation and the physical behavior of sampled-data systems is essential for maintaining theoretical rigor and self-consistency.

Some papers that use pure mathematical methods claim that their series summation results follow from the Nyquist–Shannon sampling theorem. However, the formula in Shannon's theorem contains a time-domain sinc function. This is different from the structure of the aliasing series in the DTFT. Using the Shannon theorem to justify the aliasing series is not rigorous.

Since Shannon theorem describes certain aspects of sampled-data systems, we will also discuss the connection between the Shannon's theorem, the DTFT, and sampled-data systems from a functional analysis perspective.

This work provides and proves the fix to $\mathcal{Z}$-transform, so that the corrected $\mathcal{Z}$-transform aligns with the DTFT result. With the fixed starred transform, \cite{xinli2018} will result in the exact correct model as \cite{YanNaPartI},\cite{YanNaPartII} while skipping the complicated infinite series summation evaluation. The fixed transform keeps the simplicity in mathematics form, as that of the $\mathcal{Z}$ transform. Moreover, the transform has a matrix-based Laplace-$\mathcal{Z}$ mapping process. This guarantees simplicity over traditional tabulation.

\section{Rigorous proof of the flaws of the starred transform}
\label{original_process}
\subsection{Background}
A key point in this paper is that the convolution of the impulse response with a Dirac comb must lead to the aliasing series in the frequency domain. We will give a detailed and rigorous proof of this result in this section.

In the past, some derivations of the $\mathcal{Z}$-transform started from the time domain. That approach is not rigorous. The main issue is that the impulse response often has first-kind discontinuities. The Dirac comb is a generalized function defined in the sense of distributions. There is no fully rigorous and widely accepted definition of how a distribution interacts with a point of discontinuity.

In contrast, our method works in the $\mathpzc{Laplace}$ domain. The Laplace-domain transfer function is smooth. Its convolution with a $\mathpzc{Laplace}$-domain Dirac comb is well-defined in the sense of distribution theory. This avoids the mathematical difficulties that appear in time-domain treatments.

\begin{theorem}[Discrete Time Fourier Transform]
\label{theoremdtft}
Let \( x_a(t) \) be a continuous-time signal and \( x_s(t) \) its sampled version:
\[
x_s(t) = x_a(t) \sum_{n=-\infty}^{+\infty} \delta(t - nT).
\]
Then the Fourier transform of \( x_s(t) \) is given by
\[
X_s(\omega) = \frac{1}{T} \sum_{k=-\infty}^{+\infty} X_a(\omega - k\omega_0), \quad \text{with } \omega_0 = \frac{2\pi}{T}.
\]
\label{samplingtheorem}
\end{theorem}

This expression can be extended to the Laplace domain by replacing the angular frequency variable $\omega$ with the complex variable 
$s=j\omega$, yielding the aliasing summation:

\begin{equation} X_s\left( s \right) = \sum_{k=-\infty}^{\infty} \frac{X(s - j\tfrac{2\pi}{T}k)}{T},
\label{nyquistresult}
\end{equation}
where $X_s(s)$ is the Laplace transform of the sampled signal.
The sampling behavior leads to the continuation in the frequency domain ($s\rightarrow s-j\frac{2\pi}{T}k$, $k\in \mathbf{Z}$). 

In addition to the Discrete-Time Fourier Transform (DTFT), another fundamental framework for analyzing sampled-data systems is the Nyquist–Shannon sampling theorem \cite{shannon1949}. 

%

In 1950s, E.I. Jury and John R. Ragazzinni introduced the $\mathcal{Z}$ transform to model the dynamics and stability. He proposed the famous mapping $z=e^{sT}$ that maps the continuous s-domain to the discrete $\mathcal{Z}$-domain, whose multiplication with $T_s$ is also known as the impulse-invariance method. His original work stated that
\begin{equation}
X_s\left( s \right)=X^*\left( z \right)
\label{Jury_Orig_Work}
\end{equation}
where $X^*(z)$ is the z-domain representation of the sampled signal.
However, according to C.H. Wilt's and Zadeh's work, Jury's derivation contains a mathematical flaw. As a consequence, $\eqref{Jury_Orig_Work}$ does NOT hold. Instead, Wilts and Zadeh claimed the correct answer to Jury's $X^*(z)$ is:

\begin{align}
X^*(z)
=\;\frac{1}{T}\,\sum_{k=-\infty}^{\infty}X\!\Bigl(s - j\,\tfrac{2\pi}{T}\,k\Bigr)
\;+\;\frac{x(0)^+}{2}
\label{Jury_Correct_Result}
\end{align}
which does not equal $X_s(s)$ stated in \textbf{Theorem \eqref{theoremdtft}}.

We have shown that the definition of the $\mathcal{Z}$-transform under the Definition-Centric Correction is incorrect. This is because it leads to results that do not match the DTFT aliasing series. The problem comes from an incorrect treatment of the inverse Laplace transform. A proper correction to the Laplace inversion will also fix the $\mathcal{Z}$-transform definition.

In the next part, we will prove the correct relationship between the inverse Laplace transform and the sampled result. The key idea still relies on the aliasing series given by the DTFT.

\section{Correction In The Functional Analysis Point Of View}

To facilitate analysis, we define the resolvent of the system matrix A as $\rho_A(s) = (sI-A)^{-1}$. In addition, we define the angular sampling frequency $\omega_s$ as $2\pi / T_s$.

\label{Correction}
\subsection{On the connection between discrete state space representation and aliasing summation}
\begin{theorem}[Aliasing Summation Formula]
\label{aliasingsummationspace}
For the continuous–time system $(A,B,C)$ with $D=0$, and all the eigen value of $A$ is on the left half complex plane and Corrected Impulse-Invariance
discrete model
$
A_z=e^{AT_s},\;B_z=B,\;C_z=Ce^{AT_s},\;
D_z=\tfrac12 CB,
$
one has for every $s\in\mathbb C$ with
$e^{sT_s}\notin\operatorname{spec}(A_z)$
\begin{align}
G_d\!\bigl(e^{sT_s}\bigr)
  =\frac{1}{T_s}\sum_{n=-\infty}^{\infty}
      G\bigl(s+j n\omega_s\bigr)
\end{align}
where
$G(s)=C(sI-A)^{-1}B$ and
$G_d(z)=C_z(zI-A_z)^{-1}B_z+D_z$.
\end{theorem}

\begin{proof}
\textit{Step 1: partial-fraction expansion of $G(s)$.}
By Lemma~\ref{resolventexpansion} (matrix resolvent expansion),
\begin{align}
G(s)
  &=C(sI-A)^{-1}B
   =\sum_{j=1}^{q}\sum_{r=1}^{m_j}
     \frac{R_{j,r}}{(s-\lambda_j)^{r}},
\\
R_{j,r}&:=C(A-\lambda_j I)^{r-1}P_jB.
\end{align}

\textit{Step 2: aliasing sum.}
Define
\begin{align}
S(s):=\frac{1}{T_s}\sum_{n=-\infty}^{\infty}G(s+j n\omega_s).
\end{align}
With $x=s-\lambda_j$ and Lemma~\ref{highsumlemma},
\begin{align}
S(s)
&=\sum_{j,r}\frac{(-1)^{r-1}R_{j,r}}{(r-1)!}
  \frac{d^{r-1}}{dx^{r-1}}
  \Bigl[\tfrac12+\!\!\sum_{m\ge1}e^{-mT_s x}\Bigr]\label{eq:S-start}\\
&=\frac12\sum_jR_{j,1}
  +\underbrace{%
     \sum_{j,r}\sum_{m\ge1}
     R_{j,r}\frac{(mT_s)^{r-1}}{(r-1)!}
     e^{-mT_s(s-\lambda_j)}
   }_{\beta}.
   \label{beta}
\end{align}

Because $D=0$, $\tfrac12\sum_j R_{j,1}=\tfrac12CB=D_z$.

\textit{Step 3: use $N_j^{m_j}=0$ to form matrix exponentials.}
With $N_j=(A-\lambda_j I)P_j,\;N_j^{m_j}=0$,
\begin{align}
\label{nilpotentexponential}
\sum_{r=1}^{m_j}\frac{(mT_s)^{r-1}}{(r-1)!}
  (A-\lambda_j I)^{r-1}P_j
  =e^{(A-\lambda_j I)mT_s}P_j.
\end{align}
Substitute equation \eqref{nilpotentexponential} into equation \eqref{beta}.
\begin{align}
\beta
    &=\sum_{j,r}\sum_{m\ge1} C e^{N_j mT_s}P_j B\,e^{-mT_s (s-\lambda_j)} \\
  &=\sum_{j,r}\sum_{m\ge1} C e^{A mT_s}P_j B\,e^{-mT_s s},
\end{align}
and using $\sum_jP_j=I$
\begin{align}
&S(s)=\tfrac12 CB
     +C\sum_{m=1}^{\infty}e^{A mT_s}B\,e^{-mT_s s}\\
     &=\tfrac12 CB+Ce^{A T_s}\sum_{m=1}^{\infty}e^{A (m-1)T_s}B\,e^{-mT_s s}
\end{align}

\textit{Step 4: geometric sum.}
Put $z=e^{sT_s}$ so that $e^{-mT_s s}=z^{-m}$. Then


Using the following lemma:
\begin{lemma}[Neumann Expansion of z domain state space]
\label{discretespacenuemannlemma}
    \begin{equation}
\begin{aligned}
&(zI - e^{A T_s})^{-1}
= \sum_{m=1}^{\infty}e^{A (m-1)T_s}z^{-m}
\end{aligned}
\end{equation}
\end{lemma}
For the detailed proof of this lemma, refer to Appendix \ref{neumannZ}.
\begin{align}
S(s)
  &=\tfrac12 CB
     +C_z(zI-A_z)^{-1}B_z \nonumber\\
     & = C_z(zI-A_z)^{-1}B_z+D_z\\
  &= G_d(z).
\end{align}
Replacing $z$ with $e^{sT_s}$ yields the desired identity.
\end{proof}

We prove that the aliasing series from the DTFT matches the result of the rigorously defined inverse Laplace transform. This shows that replacing sampled data with an inverse Laplace transform is mathematically sound. The theorem confirms that a $\mathcal{Z}$-domain transfer function is equivalent to the aliasing series sum. The aliasing series also matches the rigorous inverse Laplace result.

Note that this result does not match the traditional engineering definition of the inverse Laplace transform. That definition differs from the pure mathematical one. The proof uses the pure mathematical definition of inverse Laplace. Therefore the root of the $\mathcal{Z}$-transform issue lies in an incorrect definition of inverse Laplace. In the next subsection we show the problems with the engineering definition. We also show that the corrected inverse Laplace matches the discrete domain used in \textbf{Theorem \eqref{aliasingsummationspace}}.

\section{Bromwich paradox: the flaw of Laplace Inversion}

For any \textbf{proper} Laplace transfer functions (i.e., a rational function \( F(s) = \frac{N(s)}{D(s)} \) with \(\deg N(s) < \deg D(s)\)), its time-domain inverse Laplace transform is defined as:
\[
f(t) = \mathcal{L}^{-1} \left\{ F(s) \right\} (t) = \frac{1}{2\pi j} \int_{c - j\infty}^{c + j\infty} F(s) e^{st} \, ds
\]

Whenever \(f(t)\) exhibits a first-kind discontinuity, the inversion result must, by classical complex and Fourier analysis, satisfy:
\[
f(0) = \frac{1}{2} \left[ f(0^-) + f(0^+) \right]
\]

However, in engineering applications, the Laplace inverse of any transfer function that yields a nonzero response at \(t = 0^+\)—such as \(\mathcal{L}^{-1}\left\{ \frac{1}{s} \right\} = u(t)\)—is universally understood to be multiplied by the original Heaviside function:

\begin{equation}
u_e(t) =
\begin{cases}
0, & t < 0 \\
1, & t \ge 0
\end{cases}
\quad \Rightarrow \quad u_e(0) = 1
\label{eq:u_e}
\end{equation}

This convention extends beyond \( \frac{1}{s} \); it applies to \emph{all} inverse Laplace transforms that are nonzero at \(t = 0^+\). In every such case, the inverse is implicitly defined as:
\[
\mathcal{L}^{-1} \left\{ F(s) \right\} (t) = f_{\text{analytic}}(t) \cdot u_e(t)
\]
where \(f_{\text{analytic}}(t)\) is the continuous analytic form obtained from residues or inverse integral transform techniques.

This introduces a fundamental inconsistency: Fourier-type inversion theorems demand that the value at a jump discontinuity be the arithmetic mean, yet the Heaviside-based engineering definition systematically adopts the right-hand limit.

This conflict, which affects all proper Laplace-domain systems with nonzero initial responses. We named it as the \textbf{Bromwich Paradox}. It highlights a structural contradiction between engineering practice and mathematical inversion theory when discontinuities are present at \(t = 0\).
We will systematically consider this and propose its correction in Section \ref{resolvent correction}.
\subsection{On the correction to Laplace Inversion of the resolvent}
\label{resolvent correction}
In modern control theory, the inverse Laplace transform of the matrix resolvent \((pI - A)^{-1}\) is commonly understood as the matrix exponential \(e^{At}\) multiplied by the step function \(u(t)\), that is:
\begin{equation}
\mathcal{L}^{-1}\bigl[(pI - A)^{-1}\bigr](t)
\;=\;
e^{At} \cdot u(t).
\end{equation}

However, it is important to emphasize that the definition of the step function \(u(t)\) differs between the pure mathematics definition \(u_m(t)\) and the engineering practice\(u_e(t)\). In pure mathematics—particularly within distribution theory and functional analysis—it is conventional to define the value at \(t = 0\) to be
\begin{equation}
u_m(0) = \frac{1}{2}.
\end{equation}

This assignment reflects the interpretation of the Heaviside function as the distributional limit of a family of continuous approximations (e.g., sigmoid-type transitions), and is consistent with the theory of tempered distributions and Fourier–Laplace inversion via principal value or symmetric limit arguments.

In what follows, we shall rigorously prove—within the framework of functional analysis—that the Laplace inverse of \((pI - A)^{-1}\) yields the generalized function \(e^{At} u(t)\), where the correct assignment at \(t = 0\) must be \(u(0) = \frac{1}{2}\). This value arises naturally when evaluating the inverse transform along a symmetric Bromwich contour, and ensures compatibility with distributional identities and Riesz projection theory.

\subsection{t=0 case}
\begin{theorem}[Bromwich Contour Riesz Projection Theorem]
Let \(A\in\mathbb{C}^{n\times n}\) have eigenvalues \(\{p_j\}_{j=1}^q\).  For each \(j\), let
\[
P_j \;=\;\frac{1}{2\pi i}\oint_{C_j}(p I - A)^{-1}\,dp
\]
be the Riesz projection onto the generalized eigenspace of \(p_j\), where \(C_j\) is a small positively–oriented circle around \(p_j\). Fix \(c>\max_j\Re(p_j)\) and the counter–clockwise contour is defined as the union of the vertical line $L_C$ (\(\Re=c\)) and the arc $C_R^-$ ($Re^{i\theta}$, where $\theta\in[\tfrac\pi2+\theta_s,\,\tfrac{3\pi}2-\theta_s]$), as shown in Fig.\ref{fig:RieszProj}. As R approaches infinity, $\theta_s=\arcsin(\tfrac{c}{R})$ approaches zero and the range of $\theta$ becomes $(\tfrac\pi2,\,\tfrac{3\pi}2)$.

\begin{figure}[H]
  \centering
  \includegraphics[width=0.95\linewidth]{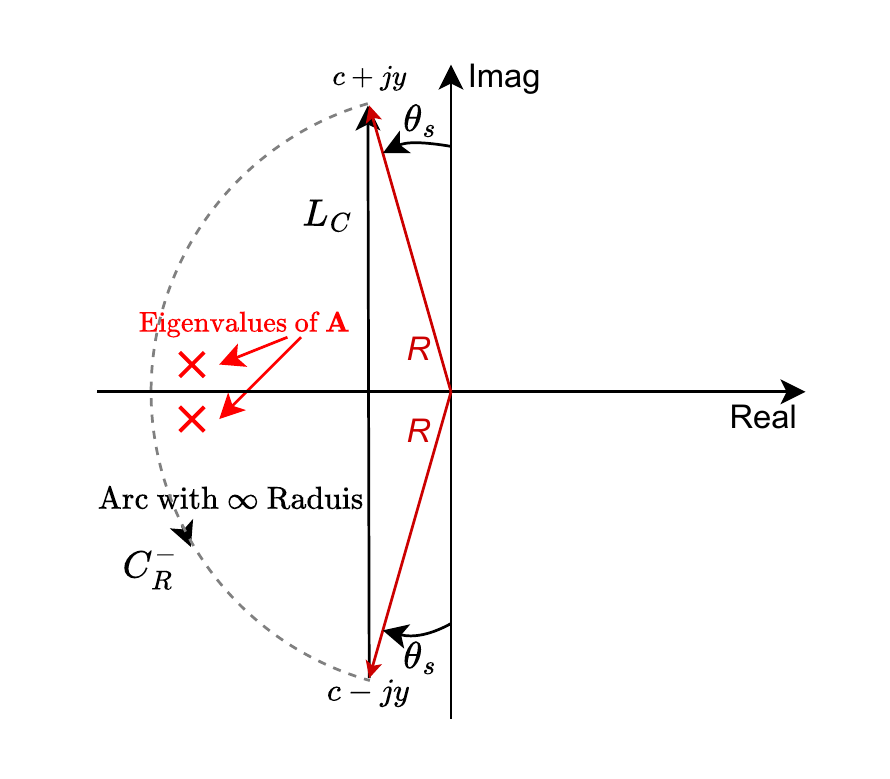}
  \caption{Integration Path of $P_j$}
  \label{fig:RieszProj}
\end{figure}

Then
\[
\sum_{j=1}^q Projection_j
=\frac{1}{2\pi i}\oint_{\mathcal{C}}(p I - A)^{-1}\,dp
=I.
\]
\end{theorem}

\begin{proof}
Since \(c>\max_j\Re(p_j)\), the closed contour \(\mathcal{C}=L_c\cup C_R^-\) encloses all eigenvalues of \(A\).  For \(|p|>c\), the resolvent has the Neumann Series expansion
\[
(p I - A)^{-1}
=\frac{1}{p}\sum_{k=0}^{\infty}\Bigl(\tfrac{A}{p}\Bigr)^k
=\sum_{k=0}^{\infty}\frac{A^k}{p^{k+1}},
\]
which converges uniformly on \(\mathcal{C}\) as \(R\to\infty\).  Hence
\[
\sum_{j=1}^q Projection_j
=\frac{1}{2\pi i}\sum_{k=0}^\infty
\oint_{\mathcal{C}}\frac{A^k}{p^{k+1}}\,dp.
\]
For \(k\ge1\), \(\|A^k\|/|p|^{k+1}=O(R^{-1})\) on the half–circle and the integrand is analytic on the line, so those integrals vanish in the limit.  Only the \(k=0\) term remains:
\[
\oint_{\mathcal{C}}\frac{dp}{p}
=\lim_{R\to\infty}\Bigl(\underbrace{\int_{L_c}\frac{dp}{p}}_{=\pi i}
+\underbrace{\int_{C_R^-}\frac{dp}{p}}_{=\pi i}\Bigr)
=2\pi i.
\]
These two parts of the integral can be calculated using the Big Arc Lemma.
Therefore
\[
\sum_{j=1}^q Projection_j
=\frac{1}{2\pi i}\,I\,(2\pi i)
=I,
\]
as claimed.

\end{proof}
Then, it is clear that:
\begin{align*}
&\frac{1}{2\pi i}\int_{L_c}(p I - A)^{-1}\,\mathrm{d}p\\
&=
\frac{1}{2\pi i}\Biggl(
\underbrace{\oint_{\mathcal{C}}(p I - A)^{-1}\,\mathrm{d}p}_{=I}
\;-\;
\underbrace{\int_{C_R^-}(p I - A)^{-1}\,\mathrm{d}p}_{=\tfrac12 I}
\Biggr)\\
&=I-\tfrac12I
=\tfrac12I.
\end{align*}

\subsection{the case for t bigger than zero}
\begin{proof}
We start from the Neumann series expansion
\[
(pI - A)^{-1}
\;=\;
\frac{1}{p}\sum_{k=0}^{\infty}\biggl(\frac{A}{p}\biggr)^{k}
\;=\;
\sum_{k=0}^{\infty}\frac{A^{k}}{p^{\,k+1}}.
\]
In the Bromwich inversion integral 
\[
\frac{1}{2\pi i}
\int_{L_{c}\cup C_{R}^{-}} e^{p\,t}\,(pI - A)^{-1}\,dp,
\]
we substitute the above series term by term:
\begin{align}
&\frac{1}{2\pi i}\int_{L_{c}\cup C_{R}^{-}} e^{p\,t}\,(pI - A)^{-1}\,dp \nonumber \\
=&
\frac{1}{2\pi i}
\sum_{k=0}^{\infty}
\int_{L_{c}\cup C_{R}^{-}}
e^{p\,t}\,\frac{A^{k}}{\,p^{\,k+1}\!}\,dp. 
\label{eq:neumann-splitting}
\end{align}

\noindent\textbf{(i) Vanishing of all terms with \(k\ge1\).}  Fix \(t>0\).  For each integer \(k\ge1\), consider 
\[
G_{k}(p)\;=\;\frac{A^{\,k}}{\,p^{\,k+1}\!}, 
\quad
p \;=\;R\,e^{\,i\theta},\;\theta\in\bigl[\tfrac{\pi}{2},\,\tfrac{3\pi}{2}\bigr].
\]
On the left half‐circle \(C_{R}^{-}\), we have \(\Re(p)\le0\), so \(\lvert e^{p\,t}\rvert = e^{t\,\Re(p)} \le1\).  Moreover,
\(\|\,A^{\,k}/p^{\,k+1}\|\le \|A\|^{k}/R^{\,k+1}\to0\) as \(R\to\infty\).  By the Arc‐Lemma (``large‐arc integral vanishes''), 
\[
\int_{C_{R}^{-}} 
G_{k}(p)\,e^{p\,t}\,dp 
\;\xrightarrow[R\to\infty]{}\;0.
\]
Meanwhile, on the vertical line \(L_{c}\) (where \(\Re(p)=c>\rho(A)\)), the integrand \(G_{k}(p)\,e^{p\,t}\) is analytic and bounded by a constant times \(e^{c\,t}\,/|p|^{\,k+1}\).  Letting \(R\to\infty\) shows 
\(\displaystyle \int_{L_{c}}G_{k}(p)\,e^{p\,t}\,dp\to0\) as well.  Hence for each \(k\ge1\),
\[
\lim_{R\to\infty}
\int_{L_{c}\cup C_{R}^{-}} e^{p\,t}\,\frac{A^{k}}{\,p^{\,k+1}\!}\,dp
\;=\;0.
\]

\noindent\textbf{(ii) The remaining \(k=0\) term.}  When \(k=0\), the integrand becomes 
\[
\frac{A^{0}}{p^{1}}\,e^{p\,t} \;=\;\frac{e^{p\,t}}{p}\;.
\]
Again the Arc‐Lemma implies 
\(\displaystyle \int_{C_{R}^{-}}\frac{e^{p\,t}}{p}\,dp\to0.\) 
Recall the Dunford–Taylor (contour) representation of the matrix exponential:
\[
e^{A\,t}
\;=\;
\frac{1}{2\pi i}
\oint_{\Gamma}\!e^{z\,t}\,\bigl(zI - A\bigr)^{-1}\,dz,
\]
where \(\Gamma\) is any counterclockwise closed contour enclosing \(\sigma(A)\).  If we now replace \(\Gamma\) by the limiting contour \(L_{c}\cup C_{R}^{-}\) and let \(R\to\infty\), the large‐arc part \(C_{R}^{-}\) vanishes for the \(k=0\) term \(\frac{e^{p\,t}}{p}\).  Consequently,
\begin{align}
e^{A\,t}
&=
\frac{1}{2\pi i}
\lim_{R\to\infty}
\int_{L_{c}\cup C_{R}^{-}} 
e^{p\,t}\,(pI - A)^{-1}\,dp \notag \\
&=
\frac{1}{2\pi i}
\int_{\,c - i\infty}^{\,c + i\infty} 
\frac{e^{p\,t}}{p}\,dp.
\end{align}

as required.  

\end{proof}

Our precise discussion of the resolvent and its inverse Laplace shows that the engineering definition misuses the Dunford–Taylor integral. At $t = 0$ the Dunford–Taylor integral becomes the Riesz projection. We need the resolvent integral over the Bromwich line. The Riesz projection uses a closed path. We handle the large arc part using the large-arc lemma.

Under the traditional rule, the inverse Laplace of the resolvent is
\[
\mathcal{L}^{-1}\{(sI - A)^{-1}\} = e^{At}\,u(t).
\]
The step function uses u(0)=1. This gives
\[
e^{At}\,u(t)\big|_{t=0} = e^{A\cdot 0} = I.
\]
Our proof shows the value at zero should be
\[
e^{At}\,u(t)\big|_{t=0} = \tfrac{1}{2}I.
\]
Hence the step function must use $u(0)=\tfrac{1}{2}$. In the $\mathcal{Z}$-transform the initial output is
\[
y[0] = C\,I\,B = C\,B.
\]
After correction it is
\[
y[0] = C\bigl(\tfrac{1}{2}I\bigr)B = \tfrac{1}{2}\,C\,B.
\]
This is exactly the algebraic mean of the left and right limits of y (0). This matches the Laplace \& Fourier Inversion theorem.
Although the Bromwich paradox has been resolved and the inverse Laplace transform is now correctly computed,  the practice of taking the value of $x(nT)$ directly from the inverse Laplace transform at that point is still not justified.


\section{A Symbolic RMCF Benchmark for the Stability Gap between Corrected and conventional z-transform}

\subsection{Plant definition in real modal canonical form}
Consider the stable underdamped second-order plant with a real zero
\begin{equation}
\begin{aligned}
&G(s)=\frac{g(s+z)}{(s+\sigma)^2+\omega^2},\qquad
\sigma>0,\ \omega>0,\ z>0,\ g>0.
\end{aligned}
\end{equation}
Define the real modal canonical form (RMCF)
\begin{equation}
\begin{aligned}
&\dot{x}(t)=Ax(t)+Bu(t),\qquad y(t)=Cx(t),\\
&A=\begin{bmatrix}-\sigma&-\omega\\ \omega&-\sigma\end{bmatrix},\quad
B=\begin{bmatrix}0\\ 1\end{bmatrix},\quad
C=\begin{bmatrix}g\kappa & g\end{bmatrix},
\end{aligned}
\end{equation}
where
\begin{equation}
\begin{aligned}
&\kappa:=\frac{\sigma-z}{\omega},\qquad z=\sigma-\kappa\omega.
\end{aligned}
\end{equation}
Then
\begin{equation}
\begin{aligned}
&C(sI-A)^{-1}B
=\frac{g(s+z)}{(s+\sigma)^2+\omega^2},\qquad
CB=g.
\end{aligned}
\end{equation}

\subsection{Corrected and right-limit impulse invariance (IRI) models}
Let $T_s>0$ be the sampling period and define
\begin{equation}
\begin{aligned}
&\alpha:=e^{-\sigma T_s}\in(0,1),\quad
\theta:=\omega T_s,\quad
c:=\cos\theta,\quad s:=\sin\theta.
\end{aligned}
\end{equation}
The matrix exponential admits the closed form
\begin{equation}
\begin{aligned}
&A_d=e^{AT_s}
=\alpha\begin{bmatrix}c&-s\\ s&c\end{bmatrix}.
\end{aligned}
\end{equation}
Following the corrected IRI state-space model used in this paper,
\begin{equation}
\begin{aligned}
&A_d=e^{AT_s},\qquad B_d=B,\qquad C_d=Ce^{AT_s},\\
&D_d=\eta\,CB=\eta g,\qquad
\eta=\tfrac12\ \text{(corrected)},\ \eta=1\ \text{(right-limit)}.
\end{aligned}
\end{equation}
Note that the two models differ only in the feedthrough $D_d$.

\subsection{Algebraic-loop elimination and the effective proportional gain}
Consider the proportional negative feedback
\begin{equation}
\begin{aligned}
&e[k]=r[k]-y[k],\qquad u[k]=K\,e[k],
\end{aligned}
\end{equation}
and the discrete plant
\begin{equation}
\begin{aligned}
&x[k+1]=A_dx[k]+B_du[k],\qquad
y[k]=C_dx[k]+D_du[k].
\end{aligned}
\end{equation}
Because $D_d\neq 0$, one has the algebraic relation
\begin{equation}
\begin{aligned}
&(1+KD_d)u[k]=K\bigl(r[k]-C_dx[k]\bigr).
\end{aligned}
\end{equation}
Hence the algebraic-loop-free but exactly equivalent implementation is
\begin{equation}
\begin{aligned}
&u[k]=K_{\mathrm{eff}}\bigl(r[k]-C_dx[k]\bigr),
K_{\mathrm{eff}}:=\frac{K}{1+KD_d}=\frac{K}{1+\eta gK}.
\end{aligned}
\end{equation}
Therefore, for regulation ($r[k]\equiv 0$), the closed-loop state matrix is
\begin{equation}
\begin{aligned}
&A_{\mathrm{cl}}=A_d-B_dK_{\mathrm{eff}}C_d.
\end{aligned}
\end{equation}

\subsection{Closed-form trace, determinant, and the stability boundary in $K_{\mathrm{eff}}$}
For the RMCF benchmark above, $C_d=CA_d$ yields
\begin{equation}
\begin{aligned}
&C_d
=\alpha\begin{bmatrix}g(\kappa c+s) & g(c-\kappa s)\end{bmatrix}.
\end{aligned}
\end{equation}
A direct computation gives
\begin{equation}
\begin{aligned}
&\mathrm{tr}(A_{\mathrm{cl}})
=2\alpha c-\alpha gK_{\mathrm{eff}}(c-\kappa s),\\
&\det(A_{\mathrm{cl}})
=\alpha^2(1-gK_{\mathrm{eff}}).
\end{aligned}
\end{equation}
Let
\begin{equation}
\begin{aligned}
&p(\lambda):=\det(\lambda I-A_{\mathrm{cl}})
=\lambda^2-\mathrm{tr}(A_{\mathrm{cl}})\lambda+\det(A_{\mathrm{cl}}).
\end{aligned}
\end{equation}
The Jury conditions for a second-order polynomial imply that stability can be monitored via
$p(\pm 1)>0$ together with $1-\det(A_{\mathrm{cl}})>0$.
For $K_{\mathrm{eff}}\ge 0$, we always have $\det(A_{\mathrm{cl}})\le \alpha^2<1$, hence
$1-\det(A_{\mathrm{cl}})>0$ automatically.
Moreover, in the practically relevant regime
\begin{equation}
\begin{aligned}
&c-\kappa s\ge \alpha,
\end{aligned}
\end{equation}
one has $p(1)>0$ for all $K_{\mathrm{eff}}\ge 0$, and the stability boundary is reached at
$p(-1)=0$, i.e.,
\begin{equation}
\begin{aligned}
&p(-1)=1+\mathrm{tr}(A_{\mathrm{cl}})+\det(A_{\mathrm{cl}})=0.
\end{aligned}
\end{equation}
Solving for $K_{\mathrm{eff}}$ yields the closed-form critical value
\begin{equation}
\begin{aligned}
&K_{\mathrm{eff}}^\star
=\frac{1+2\alpha c+\alpha^2}{g\bigl(\alpha^2+\alpha(c-\kappa s)\bigr)}.
\end{aligned}
\end{equation}

\subsection{Mapping back to the real gain $K$ and a sharp condition for a dramatic stability gap}
The real proportional gain $K$ and the effective gain $K_{\mathrm{eff}}$ satisfy
\begin{equation}
\begin{aligned}
&K_{\mathrm{eff}}=\frac{K}{1+\eta gK},
K=\frac{K_{\mathrm{eff}}}{1-\eta gK_{\mathrm{eff}}},\ \ (\eta gK_{\mathrm{eff}}<1).
\end{aligned}
\end{equation}
Hence the (positive) stability region in $K$ is
\begin{equation}
\begin{aligned}
&0<K<K_{\max}(\eta),
K_{\max}(\eta)=\begin{cases}
\dfrac{K_{\mathrm{eff}}^\star}{1-\eta gK_{\mathrm{eff}}^\star}, & \eta gK_{\mathrm{eff}}^\star<1,\\[6pt]
+\infty, & \eta gK_{\mathrm{eff}}^\star\ge 1.
\end{cases}
\end{aligned}
\end{equation}
Since $\lim_{K\to+\infty}K_{\mathrm{eff}}=1/(\eta g)$, the corrected model ($\eta=\tfrac12$)
saturates at $2/g$, while the right-limit model ($\eta=1$) saturates at $1/g$.
Therefore, a dramatic stability gap occurs when
\begin{equation}
\begin{aligned}
&\frac{1}{g}\le K_{\mathrm{eff}}^\star<\frac{2}{g},
\end{aligned}
\end{equation}
under which the right-limit model predicts $K_{\max}(1)=+\infty$ (no upper bound),
whereas the corrected model yields a finite upper bound
$K_{\max}(\tfrac12)=K_{\mathrm{eff}}^\star/(1-\tfrac12 gK_{\mathrm{eff}}^\star)$.

\section{Conclusion}
A conclusion section is not required. Although a conclusion may review the 
The main points of the paper, do not replicate the abstract as the conclusion. A 
conclusion might elaborate on the importance of the work or suggest 
applications and extensions.

\appendices
\section{Lemma 1: Classical Cotangent Identity}

\[
\sum_{n=-\infty}^{\infty} \frac{1}{x + n} \;=\; \pi \,\cot(\pi x),
\quad x \notin \mathbb{Z}.
\]

\section{Lemma 2: Half-Part Expansion}

\[
\frac{1}{1 - e^{-sT_s}} \;-\; \frac{1}{2}
\;=\;
\coth\!\Bigl(\tfrac{sT_s}{2}\Bigr)
\]

\section{Lemma4:state space representation of forwarded state transfer function}
\label{state space no-delay}
Let $(A_d, B_d, C_d, D_d)$ be a discrete-time state-space realization of a system, and suppose that $D_d = 0$. Then the transfer function
\[
G(z) = z\, C_d (z I - A_d)^{-1} B_d
\]
admits a realization $(A_{\mathrm{new}}, B_{\mathrm{new}}, C_{\mathrm{new}}, D_{\mathrm{new}})$ of the form
\begin{align*}
A_{\mathrm{new}} &= A_d, \\
B_{\mathrm{new}} &= B_d, \\
C_{\mathrm{new}} &= C_d A_d, \\
D_{\mathrm{new}} &= C_d B_d.
\end{align*}

\subsection{Key Algebraic Identity}

To express \( G_{\text{new}}(z) \) in the form of a standard state-space transfer function (i.e., without explicitly appearing \( z \)-factors and instead represented as \((zI - A)^{-1}\)), we use the following key matrix identity:
\[
z (zI - A_d)^{-1} = I + A_d (zI - A_d)^{-1}.
\]

\section{Big Arc Lemma}
\label{section:big_arc_lemma}
\begin{lemma}[Large Arc Lemma]
Let $f(z)$ be a complex function analytic in a neighborhood of infinity.
Assume that as $|z|\to\infty$, $z f(z)$ tends uniformly to a constant $K$ 
in the sector $\theta_1 \leq \arg z \leq \theta_2$. Then
\begin{align}
\lim_{R \to \infty} \int_{C(R)} f(z)\,dz
= i(\theta_2 - \theta_1)\,K,
\end{align}
where $C(R)$ is a counterclockwise circular arc of radius $R$, centered at the origin,
and spanning the angle range $\theta_1 \leq \arg z \leq \theta_2$.
\end{lemma}
\section{Kernal Expansion Lemma}
\begin{lemma}
Let $T_s>0$ and set $\omega_s=2\pi/T_s$.  Then for all complex
$\sigma\notin -j\,\omega_s\mathbb Z$,
\begin{align}
\frac{1}{e^{\sigma T_s}-1}
&=
\frac{1}{T_s}\sum_{n=-\infty}^{\infty}\frac{1}{\sigma+j\,n\,\omega_s}
-\frac12.
\end{align}
\end{lemma}

\section{Expansion Of Resolvent Lemma}

\begin{lemma}[Partial Fraction Expansion of the Resolvent] \cite{dunford1988linear}
Let \( A \in \mathbb{C}^{n \times n} \) be a square matrix with distinct eigenvalues \( \{\lambda_j\}_{j=1}^{q} \), and let \( m_j \) denote the size of the largest Jordan block associated with \( \lambda_j \). Then the resolvent \( (sI - A)^{-1} \) can be expanded as:
\begin{align}
(sI - A)^{-1}
= \sum_{j=1}^{q} \sum_{r=1}^{m_j}
  \frac{(A - \lambda_j I)^{r - 1} P_j}{(s - \lambda_j)^r}, \label{eq:resolvent_expansion}
\end{align}
where \( P_j \) is the Riesz projection onto the generalized eigenspace associated with \( \lambda_j \).
\label{resolventexpansion}
\end{lemma}
\section{M order kernal derivation}
\begin{lemma}[Higher‐order Partial Fractions via Differentiation]
\label{highsumlemma}
Let \(T_s>0\) be the sampling period and \(\omega_s = 2\pi/T_s\).  Then for every integer \(r\ge1\) and for all \(s\) with \(\Re(s)>0\), one has
\begin{align}
&\frac{1}{T_s}\sum_{n=-\infty}^{\infty}\frac{1}{\bigl(s + jn\omega_s\bigr)^{r}}
\;\\&=\;
\frac{(-1)^{\,r-1}}{(r-1)!}\,
\frac{d^{\,r-1}}{ds^{\,r-1}}
\Biggl[\frac12 \;+\;\sum_{m=1}^{\infty}e^{-mT_s s}\Biggr].
\label{eq:lemma_high_order}
\end{align}
\end{lemma}

\section{Non-delay Poisson Summation}

\begin{lemma}[Poisson Summation—Zero-Phase Case]

\label{poissonsum0}
Let \(f \in \mathscr{S}(\mathbb{R})\) be a Schwartz function with Fourier transform
\[
\mathcal{F}(\omega) \;=\; \int_{-\infty}^{\infty} f(t)\,e^{-i\omega t}\,dt.
\]
Then
\[
\boxed{
\sum_{n=-\infty}^{\infty} f(n)
\;=\;
\sum_{k=-\infty}^{\infty} \mathcal{F}(2\pi k).
}
\]
\end{lemma}


As a special case, set \(x = 0\) in the shifted Poisson formula
\[
\sum_{n=-\infty}^{\infty} f(x + n)
\;=\;
\sum_{k=-\infty}^{\infty} F(2\pi k) \, e^{2\pi i k x}.
\]
This immediately gives
\[
\sum_{n=-\infty}^{\infty} f(n)
\;=\;
\sum_{k=-\infty}^{\infty} F(2\pi k).
\]
In other words, after computing the continuous Fourier transform \(F(\omega)\),
replace \(\omega\) by \(2\pi k\) and sum over all integers \(k\) to recover the
time-domain point-value sum, exactly as stated in
\textbf{Theorem~\ref{poissonsum0}}.

\section{Discrete State Space summation Lemma}
\label{neumannZ}
\begin{equation}
\begin{aligned}
&(zI - e^{A T_s})^{-1}
= z^{-1}\,(I - e^{A T_s}z^{-1})^{-1}
\\
&= z^{-1}\,(I - e^{A T_s}e^{-sT_s})^{-1}
   \quad\bigl(z=e^{sT_s}\bigr)
\\
&= z^{-1}\,(I - e^{(A - sI)T_s})^{-1}
= z^{-1}\sum_{k=0}^{\infty}\!\bigl(e^{(A - sI)T_s}\bigr)^{k}\\
   &\quad\text{(By Neumann series)}
\\
&= z^{-1}\sum_{k=0}^{\infty}e^{A k T_s}e^{-sT_s\,k}
= z^{-1}\sum_{k=0}^{\infty}e^{A k T_s}z^{-k}
\\
&= \sum_{k=1}^{\infty}e^{A (k-1)T_s}z^{-k}.
\end{aligned}
\end{equation}

\section*{References}

\bibliographystyle{IEEEtran}  
\bibliography{reference}     
\begin{IEEEbiography}[{\includegraphics[width=1in,height=1.25in,clip,keepaspectratio]{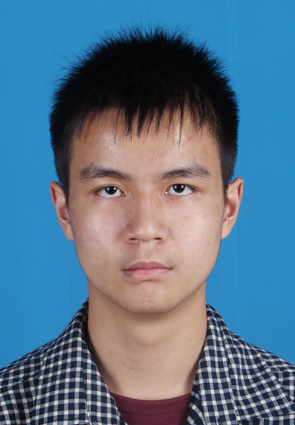}}]{Yuxin Yang} (Member, IEEE) was born in China. He received the Bachelor’s degree in electrical engineering from the University of New South Wales (UNSW), Sydney, Australia, in 2024. He is currently pursuing the Master by Philosophy degree in electrical engineering at UNSW. His major field of research is power electronics and control systems.

He is currently conducting research on small-signal modeling of power electronic systems, sampled-data control theory, and the foundational theory of control systems. His current research focuses on unifying sampled-data models with continuous-time system representations using rigorous mathematical tools.
\end{IEEEbiography}

\begin{IEEEbiography}[{\includegraphics[width=1in,height=1.25in,clip,keepaspectratio]{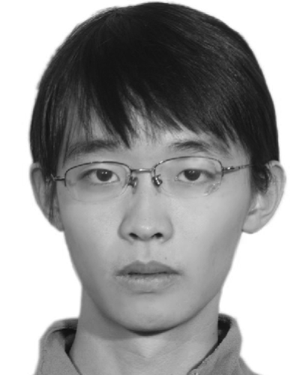}}]{Hang Zhou} (Member, IEEE) received the bachelor’s and Ph.D. degrees in electrical engineering from the University of New South Wales, Sydney, NSW, Australia, in 2017 and 2022, respectively.

He is currently a Research Associate with the University of New South Wales in the field of power electronics, specializing in dc–dc converters. His research interests include low-voltage high-current dc–dc topologies, small-signal modeling of dc–dc converters, three-phase power factor correction, and high-voltage power sources.
\end{IEEEbiography}

\begin{IEEEbiography}[{\includegraphics[width=1in,height=1.25in,clip,keepaspectratio]{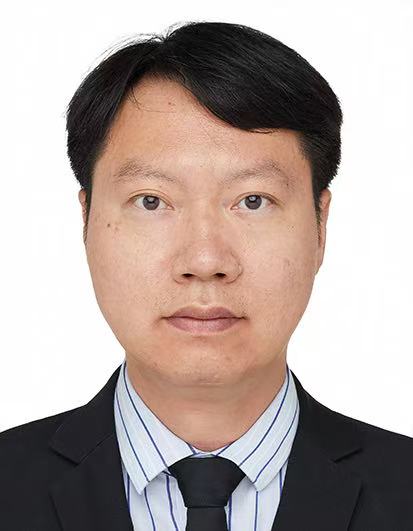}}]{Chaojie Li} (Member, IEEE) received the B.Eng. degree in electronic science and technology and the M.Eng. degree in computer science from Chongqing University, Chongqing, China, in 2007 and 2011, respectively, and the Ph.D. degree from RMIT University, Melbourne, Australia, in 2017, where he was a Research Fellow for one and a half years. He was a Senior Algorithm Engineer at Alibaba Group. He is a Senior Research Associate with the School of Electrical Engineering and Telecommunications, the University of New South Wales (UNSW). He was a recipient of the ARC Discovery Early Career Researcher Award in 2020. His current research interests include graph representation learning, distributed optimization and control in smart grid, neural networks, and their application.
\end{IEEEbiography}
\begin{IEEEbiography}[{\includegraphics[width=1in,height=1.25in,clip,keepaspectratio]{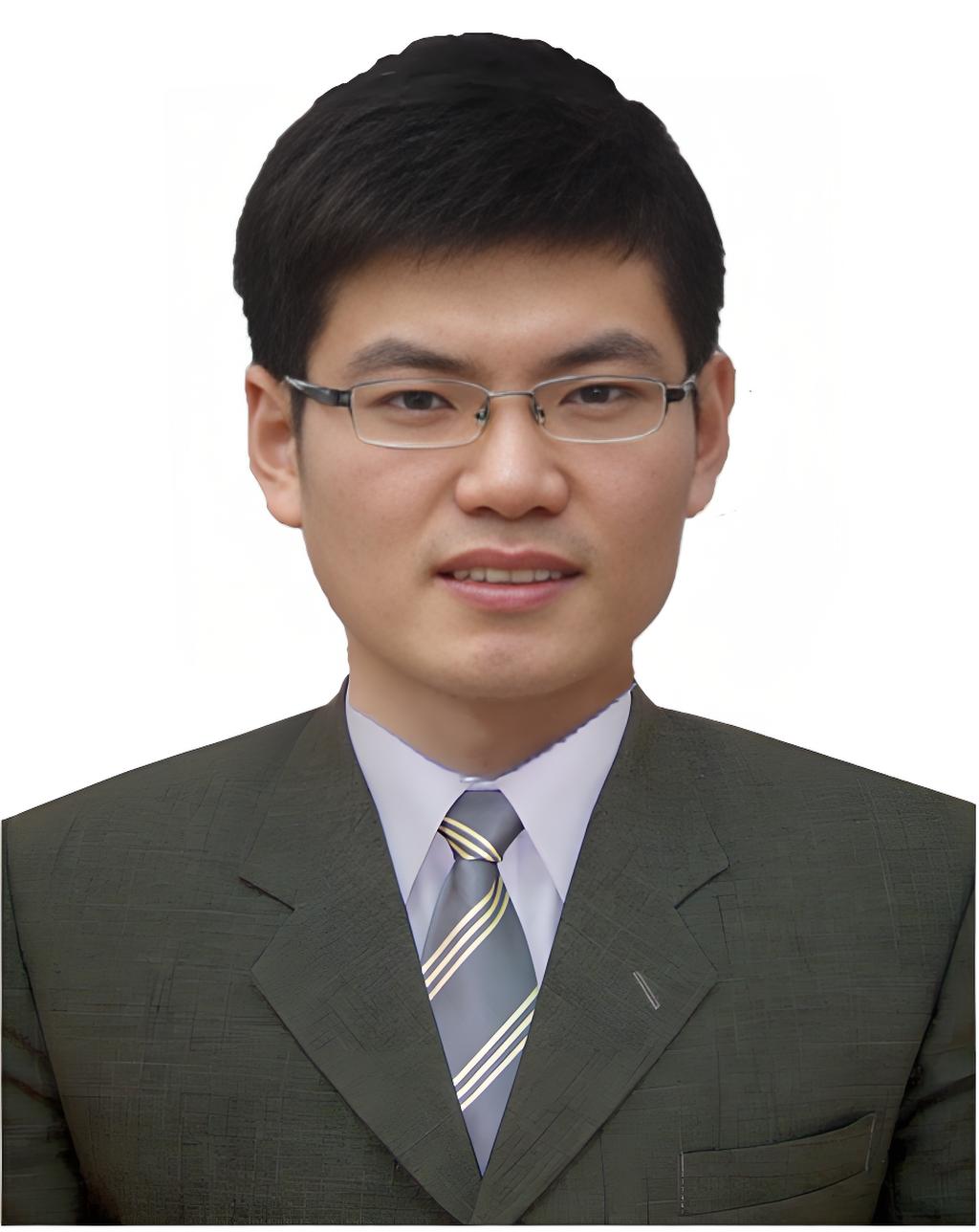}}]{Xin Li} (Member, IEEE) received the B.S. and Ph.D. degrees in electrical engineering and automation from Nanjing University of Aeronautics and Astronautics, Nanjing, China, in 2012 and 2018, respectively.,In 2019, he was a Research Engineer with Huawei Technologies Company Ltd., Shanghai, China. From 2020 to 2022, he was a Research Fellow with Nanyang Technological University, Singapore. Since Aug. 2022, has been an Associate Researcher with the School of Electrical Engineering, Southeast University, Nanjing, China. His current research interests include modeling, control and design of PWM converter, resonant converter, and wireless power transfer system.
\end{IEEEbiography}
\begin{IEEEbiography}[{\includegraphics[width=1in,height=1.25in,clip,keepaspectratio]{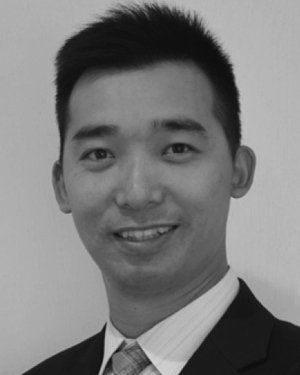}}]{Yingyi Yan} (Member, IEEE) received the bachelor’s degree from Zhejiang University, Hangzhou, China, in 2007, and the M.S. and Ph.D. degrees from the Center for Power Electronics Systems, Virginia Tech, Blacksburg, VA, USA, in 2010 and 2013, respectively.,He was an Application Engineer with Linear Technology (now part of Analog Devices), Milpitas, CA, USA, where he is currently a Senior IC Design Engineer for power products. Since 2013, he has been with Linear Technology (now part of Analog Devices). He is responsible for developing controllers for point-of-load (PoL) applications, intermediate-bus power converters, and high-current integrated power stages. He holds eight U.S. patents. His research interests include modeling and analysis of power converters, advanced topologies and control techniques, and high-frequency power conversion.
\end{IEEEbiography}
\begin{IEEEbiography}[{\includegraphics[width=1in,height=1.25in,clip,keepaspectratio]{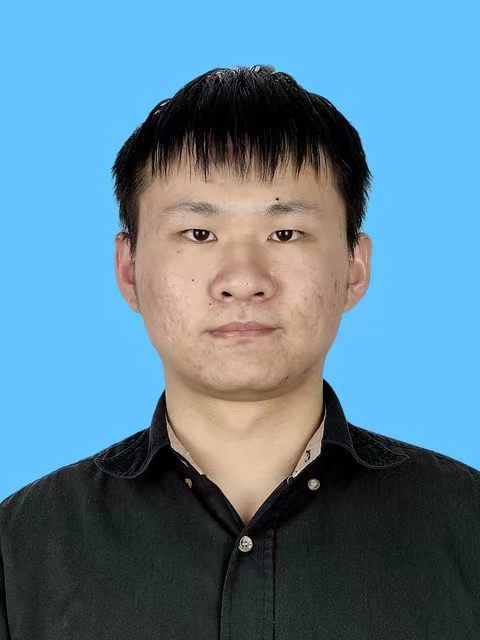}}]{Mingyang Zheng}
received the B.Eng. degree in electrical engineering from South China University of Technology in 2025. He is currently with Incosync Co., Ltd., where he works as a hardware engineer. His research interests include analog circuit design and power electronics.
\end{IEEEbiography}

\end{document}